# Construct Particle-Packed Configuration Enable Excellent Thermal Insulation Performance

Zizhen Lin[1,2], Congliang Huang[1*]

[1]School of Electrical and Power Engineering, China University of Mining and Technology, Xuzhou 221116, China.

[2]Key Laboratory of Thermo-Fluid Science and Engineering of MOE, School of Energy and Power Engineering, Xi'an Jiaotong University, Xi'an 710049, China.

a) Author to whom correspondence should be addressed. E-mail: huangcl@cumt.edu.cn.

**Abstract:** The nanoparticle packed bed (NPB) as a kind of promising thermal insulation materials has drawn widely concern because of their quite low thermal conductivities ($k$). In this paper, based on the concept that NPB morphology enable low $k$, we further proposed a common method, a hybrid strategy, to further reduce $k$ and enhance mechanical strength, simultaneously. The lowest effective thermal conductivity ($k_e$) of hybrid NPB can even as low as 0.018Wm⁻¹K⁻¹, which is much lower than the $k$ of the free air and most common thermal insulation materials, due to the quite low solid-phase thermal conductivity ($k_s$), negligible thermal conductivity of the confined air ($k_a$) and small radiative thermal conductivity ($k_r$). Neglecting the $k_a$, the minimum $k_e$ occurs at the porosity where domination role changes from $k_s$ to $k_r$. In addition, an excellent mechanical strength, nearly 40-50 % compared to bulk silica, was also harvested in silica hybrid NPB. This study is expected to supply some information for thermal insulation material design.

---
*Corresponding author.
E-mail: huang198564@gmail.com (C.-L. Huang).

## 1. Introduction

The silica nanoporous materials, such as silica aerogel [1-3], MCM-41 and SBA-15 [4,5], have drawn a wide interest because of their quite low thermal conductivities ($k$) for thermal insulation applications. Even without considering the influence of air and the radiation in pores of a nanoporous material, the thermal conductivity of the solid phase should still be larger than the Einstein limit ($k_E$) [6] where $k_E$ is the $k$ of the corresponding amorphous bulk. Even owning the lowest $k_E$, a thermal insulation material is still somewhat not competent for its applications in several areas, such as high temperature energy storage tanks [7], gas turbine engines and space applications [8,9]. Thus, to probe a nanoporous material with a quite low thermal conductivity is still desirable.

Compared to the traditional nanoporous material, the nanoparticle packed bed (NPB) as one kind of typical powder moulding material possesses a high porosity with air as matrix and nanoparticles as fillers, and thus have advantage in high density of inter-nanoparticle contact interface, which can provide an effective scattering mechanism for mid-and long-wavelength phonons that contribute heavily to the thermal conductivity ($k$). [10-13] Due to the low $k$, the NPB has excellent potential in the thermal insulation area.

There were already some researches probing the influence factors of the effective thermal conductivity ($k_e$) of NPBs for finding a proper way to reduce it. Prasher [14] theoretically investigated the $k_e$ of a NPB and reported that the contact interface between nanoparticles could effectively limit the phonon transport and therefore lead to an ultralow $k_e$. Subsequently, Hu et al. [15] confirmed the ultralow $k_e$ of NPBs with experimental works. Muftah et al. [16,17] studied the $k_e$ of copper and nickel NPB at vacuum environment and under influence of moisture respectively, and the lowest $k_e$ of them can even attend to 0.018 W m$^{-1}$ K$^{-1}$ because of the high thermal resistance between nanoparticles. In our previous works, we also confirmed that there is an important effect of the contact interface and the porosity ($\varphi$) on the $k_e$. [18,19] To our knowledge, using the hybrid nanoparticles to reduce the $k_e$ of NPB has not been reported before, when the hybrid nanoparticles could have an important effect on the

$k_e$. This work is expected to further decrease the $k_e$ of NPBs through tuning the hybrid ratio of the binary-mixed nanoparticles.

In this paper, the NPB made of hybrid silica nanoparticles (h-NPB) with diameters of 10 and 50 nm is firstly prepared with a cold-press method in which a force is utilized to press the nanoparticle powders into a tablet, and then the $k_e$ is experimentally studied. The $k_e$ of pure NPB (p-NPB) is also measured for comparison. The lowest $k_e$ of silica h-NPB is 0.018 Wm$^{-1}$K$^{-1}$ at a porosity ($\varphi$) of 0.53, which is much lower than that of p-NPB measured in this work and also the ones reported in previous works [20]. To reveal the underlying mechanism of the lowest $k_e$, the contribution of the air thermal conductivity ($k_a$), the radiative thermal conductivity ($k_r$) and the solid phase thermal conductivity ($k_s$) are studied respectively. This study is expected to supply some information for thermal insulation material design.

## 2. Sample preparation and characterizations

### 2.1 Nanoparticle powder characterization

The silica nanoparticles with diameters of 10 nm and 50 nm are commercially obtained from Beijing DK Nano Technology Co., Ltd. The transmission electron microscope (TEM) method is applied to observe the microstructures of nanoparticles by Tecnai G2 F20 instrument (FEI Co., USA) with an accelerating voltage of 80 kV. Results are shown in Fig. 1. It shows that sizes of nanoparticles are almost uniform. To investigate the effect of the hybrid ratio of nanoparticles on the $k_e$, two different proportions of 10-nm and 50-nm silica nanoparticles are selected, such as, 3:7 and 5:5. Hereafter, for convenience, the 50+10 nm (3:7) and the 50+10 nm (5:5) silica h-NPBs are signified as (3:7) and (5:5) NPBs for convenience.

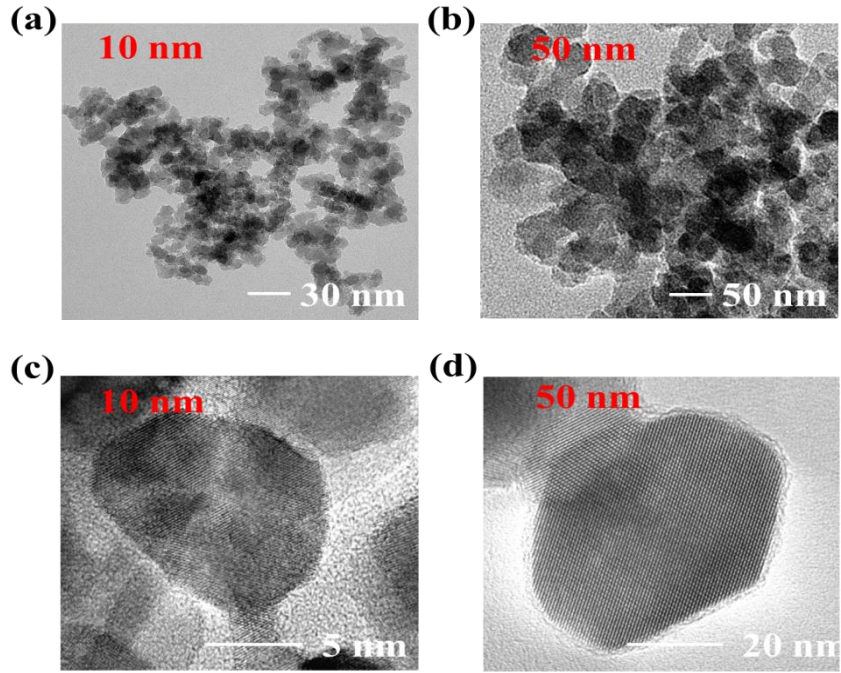

*Fig. 1 Microstructures of silica nanoparticles: (a) 10 nm silica nanoparticles obtained by TEM; (b) 50 nm silica nanoparticles obtained by TEM; (c) 10 nm silica nanoparticles obtained by HRTEM; (d) 50 nm silica nanoparticles obtained by HRTEM.*

## 2.2  NPB preparation and structure characterization

After treating the hybrid nanoparticles with a physical ultrasonic dispersion for 60 minutes, a ball-milling method is utilized to blend them (200 revolutions per minute for 60 minutes, then 100 revolutions per minute for 60 minutes). Finally, the NPBs are prepared with the cold-pressing method. [19, 21] The preparation process is illustrated in Fig. 2 (a). Sixteen different stamping pressures (1, 2, 5, 10, 12, 14, 16, 18, 20, 22, 24, 26, 30, 34, 40 and 46 MPa) are applied to obtain different-porosity samples. The (3:7) NPB prepared with a pressure of 20 MPa is shown in Fig.2 (b), and its microstructure is shown in Fig. 2 (c), which is observed with the field-emission scanning electron microscopy (SEM) on a QuantaTM-250 instrument (FEI Co., USA) with an accelerating voltage of 30 kV. It shows that there is no obvious deformation of nanoparticles, and the hybrid nanoparticles are uniformly mixed. Microstructure of (3:7) NPB prepared with a pressure of 40 MPa is also characterized for comparison as shown in Fig. 2 (d), the nanoparticle aggregation is widely appearing in view, and an obvious nanoparticle deformation is observed.

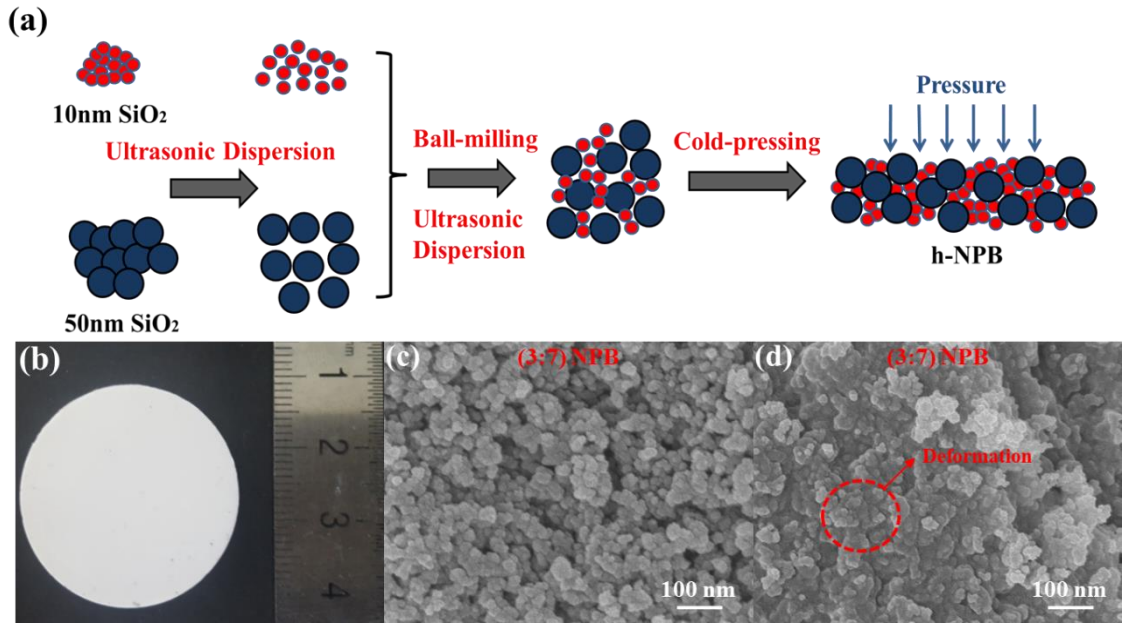

*Fig. 2 Preparation, macrostructure and micrestructure characteriation of h-NPB: (a) Preparation process of h-NPB; (b) Macrostructure of h-NPB; (c) Micrestructure of (3:7) h-NPB, which prepared with a pressure of 20 MPa, observed by the SEM; (d) Micrestructure of (3:7) h-NPB, which prepared with a pressure of 40 MPa, observed by the SEM*

## 2.3 Thermal conductivity and hardness measurement

The $k_e$ of NPBs is measured with the hot-wire method [22, 23] using a commercial device (Model TC3000, Xian XIATECH Technology Co.). The hot-wire method equipment is shown in Fig. 3(a), and more details about the $k_e$ measurement can be found in our previous works [11, 18]. With every sample measured along five different directions, a mean $k_e$ is obtained with a deviation less than 4.5%. The $k_e$ is also measured by the transient heat source method (TPS), as shown in Fig. 3(b), for comparison. With every sample measured along five different directions, a mean $k_e$ is obtained with a deviation less than 7.5%. Details about the $k_e$ measurement can also be found in Ref. 11.

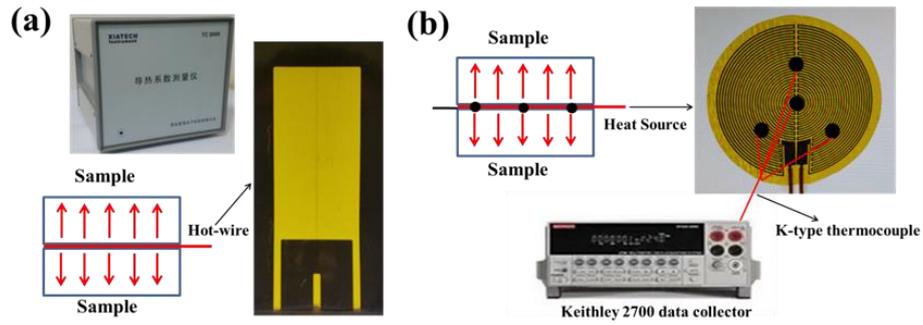

*Fig. 1. Thermal conductivity measurement: (a) hot-wire method; (b) transient plane heat source method.*

The contribution of air on the $k_e$ of a NPB is estimated by subtracting $k_e$ measured under atmospheric pressure with the one measured in a vacuum environment ($k_v$). The radiative thermal conductivity ($k_r$) is calculated by the diffusion approximation model based on the measurement of spectral transmittance ($\tau_\lambda$) observed with a Fourier transform infrared spectrometer (VERTEX 80v Germany Bruker). Details are added in Appendix A. Considering the negligible contribution of air on the $k_e$ (see section 3.2), the $k_s$ in a NPB can be calculated by $k_s = k_e - k_r$. All the thermal conductivity measurement is implemented at room temperature (300K).

As thermal insulation materials, the h-NPBs should have a high structure strength, in this work, the hardness of h-NPB is also probed. A Tukon™ 1102/1202 Vickers hardness tester is applied to measure the HV hardness. In the experiment, a force of 9.8 Newton is applied to press the diamond probe into the samples, and the force is maintained for 15 seconds. The HV of the materials can be reflected by the size of the surface indentation. To obtain the average hardness, five different locations on each NPB are selected to measure the HV. The relative error of the HV is less than 8.3%.

## 3. Results and Discussions

Firstly, the $k_e$ and HV of p-NPB and h-NPB is discussed in section 3.1. Several popular silica-based thermal insulation materials are also given in section 3.1 to evaluate the thermal insulation performance of h-NPBs. Secondly, the $k_a$, $k_r$ and $k_s$ are experimentally investigated to illustrate different contributions in the ultra-low $k_e$ of h-NPB in section 3.2. Finally, the underlying mechanism of the ultra-low $k_e$ of h-NPB is revealed in Part 3.3.

### 3.1 Thermal conductivity and hardness of NPBs

The $k_e$ of p-NPBs and h-NPBs are respectively shown in Fig. 4(a) and (b). The relative error of $k_e$ measured by the hot-wire method is less than 4.86% compared to that got by the transient heat source method, which confirms the reliability of $k_e$ measurement in this work. As shown in Fig. 4, the $k_e$ of p-NPBs firstly decreases and then rises with the increase of $\varphi$, which is similar to that observed in our previous works [11, 18, 20]. There is a similar changing tendency of the h-NPBs as that of the p-NPBs as show in Fig. 4(b). The $k_e$ of h-NPBs decreases gradually and reaches a minimum value of 0.018 Wm$^{-1}$K$^{-1}$ at $\varphi$=0.53, which is much lower than that of free air (about 0.026 W m$^{-1}$ K$^{-1}$) and also smaller than that of p-NPBs shown in Fig. 4(a). This ultra-low $k_e$ of h-NPB is even close to that of the silica aerogel and also less than other silica-based thermal insulation materials, such as silica-based mesoporous materials (MCM-41 and SBA-15), silica-based ceramic and 2-dimensional silica films as listed in Table 1. To unveil the underlying mechanism of the quite low $k_e$ of p-NPB and h-NPB, the contributions of $k_a$, $k_r$ and $k_s$ on $k_e$ are respectively discussed in the next part.

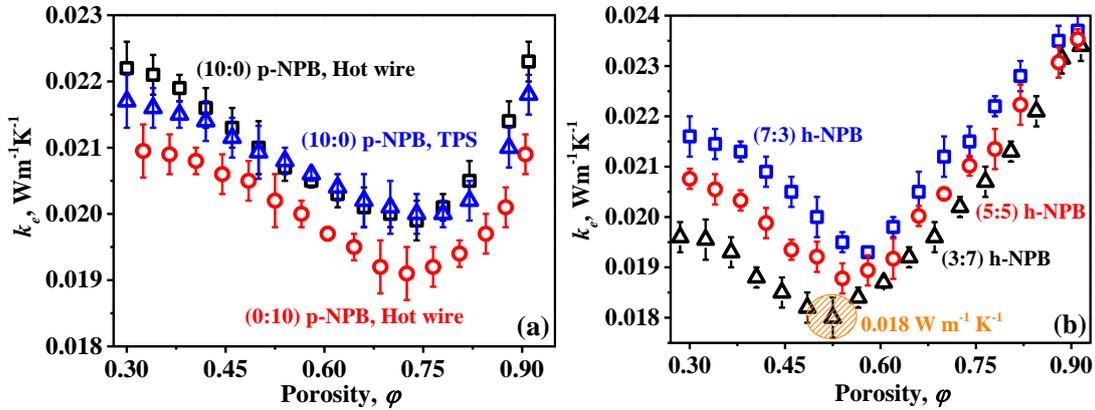

*Fig. 4 $k_e$ versus $\varphi$: (a) p-NPBs; (b) h-NPBs.*

The Vickers hardness is also measured to evaluate their mechanical properties as shown in Fig. 5. The Vickers hardness of a NPB can be about 61.7% that of the bulk silica (1167 for silica bulk), which is stronger than most popular thermal insulation materials, including silica aerogels and 2-dimensional silica films. Considering the light, high strength and low $k_e$, NPB shows a good potential in thermal insulation application.

Table 1 Thermal conductivity of different silica materials

| Materials | Scale and porosity | Thermal conductivity (Wm$^{-1}$ K$^{-1}$) | Refs. |
|---|---|---|---|
| Silica bulk | / | 1.47 | [24] |
| Silica films | 100-nm thickness | 1.30 | [24] |
| Ceramic | Macroscale, $\varphi = 0.4 - 0.5$ | 1.5-1.9 | [25] |
| Silica aerogel | $\varphi = 95\% - 97\%$ | 0.014-0.04 | [27,26] |
| MCM-41 | $\varphi = 0.71$, $d_{pore} = 3.9\ nm$ | 0.09 | [5] |
| SBA-15 | $d_{pore} = 7.83\ nm$ | 0.164 | [错误!未定义书签。] |

$d_{pore}$ is the diameter of pores.

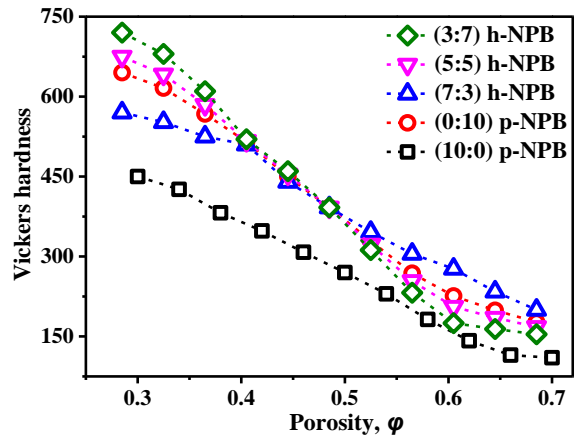

*Fig. 5 Vickers hardness of NPBs versus porosity*

## 3.2 Air thermal conductivity ($k_a$), radiative thermal conductivity ($k_r$) and solid phase thermal conductivity ($k_s$)

The $k$ of air in NPB is calculated by $k_a = k_e - k_v$, where $k_v$ is the $k$ of p-NPB and h-NPB measured in a vacuum environment, as show in Figs. 6 (a) and 6(b). The $k_a$ of p-NPB and h-NPB is shown in Figs. 6 (c) and 6(d). It shows that the $k_a$ can only take 3-5% in $k_e$, which is small enough to be neglected. Thus, we ignore the effect of $k_a$ on $k_e$ in the following up discussions.

The $k_r$ of p-NPB and h-NPB are shown in Figs. 6 (e) and 6(f), and the enhanced $k_r$ is observed for both p-NPB and h-NPB with increasing $\varphi$, which is similar to that reported in Ref. [27] for nanoporous materials. It is worth noting that the $k_r$ of p-NPBs is larger than that of h-NPBs in the whole porosity span, which can be explained by that there is a larger surface area in h-NPBs than in p-NPBs when surface will inhibit radiation heat transfer (The characterization of surface area of p-NPB and h-NPB will be discussed in section 3.3). The $k_r$ will dominate the $k_e$ of p-NPBs (h-NPB) at $0.53 \leq \varphi \leq 0.9$ $(0.72 \leq \varphi \leq 0.9)$, and this will be carefully discussed in section 3.3.

Ignoring the influence of $k_a$ on $k_e$, the $k_s$ can be experimentally obtained by $k_s = k_e - k_r$. Results are shown in Figs. 6 (g) and 6 (h). The $k_s$ will decrease with the increasing $\varphi$. The enhanced thermal resistance caused by the scattering of diffusive phonon is responsible for the reduced $k_s$ when the $\varphi$ is increased [28]. The $k_s$ will dominate the $k_e$ of p-NPBs (h-NPB) at $0.3 \leq \varphi \leq 0.53$ $(0.3 \leq \varphi \leq 0.72)$. Based on the measurement of $k_a$, $k_r$ and $k_s$, the underlying mechanism of the ultra-low $k_e$ of h-NPB will be unveiled in section 3.3.

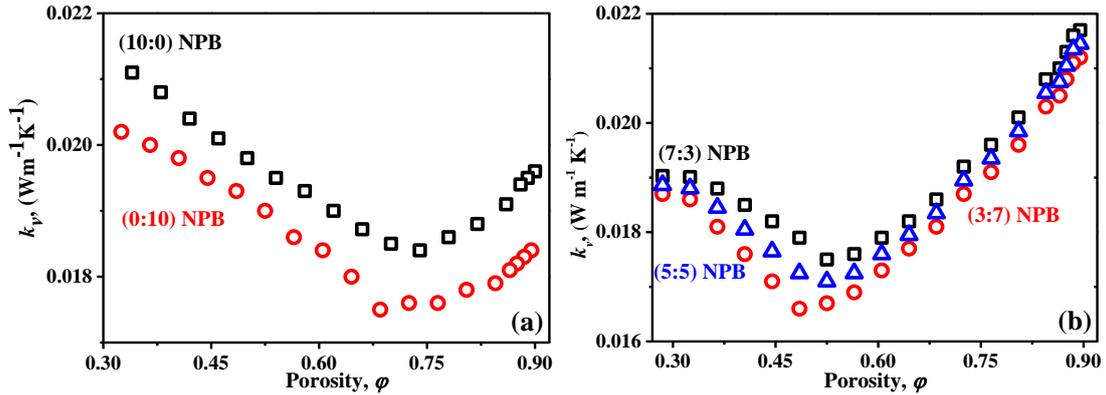

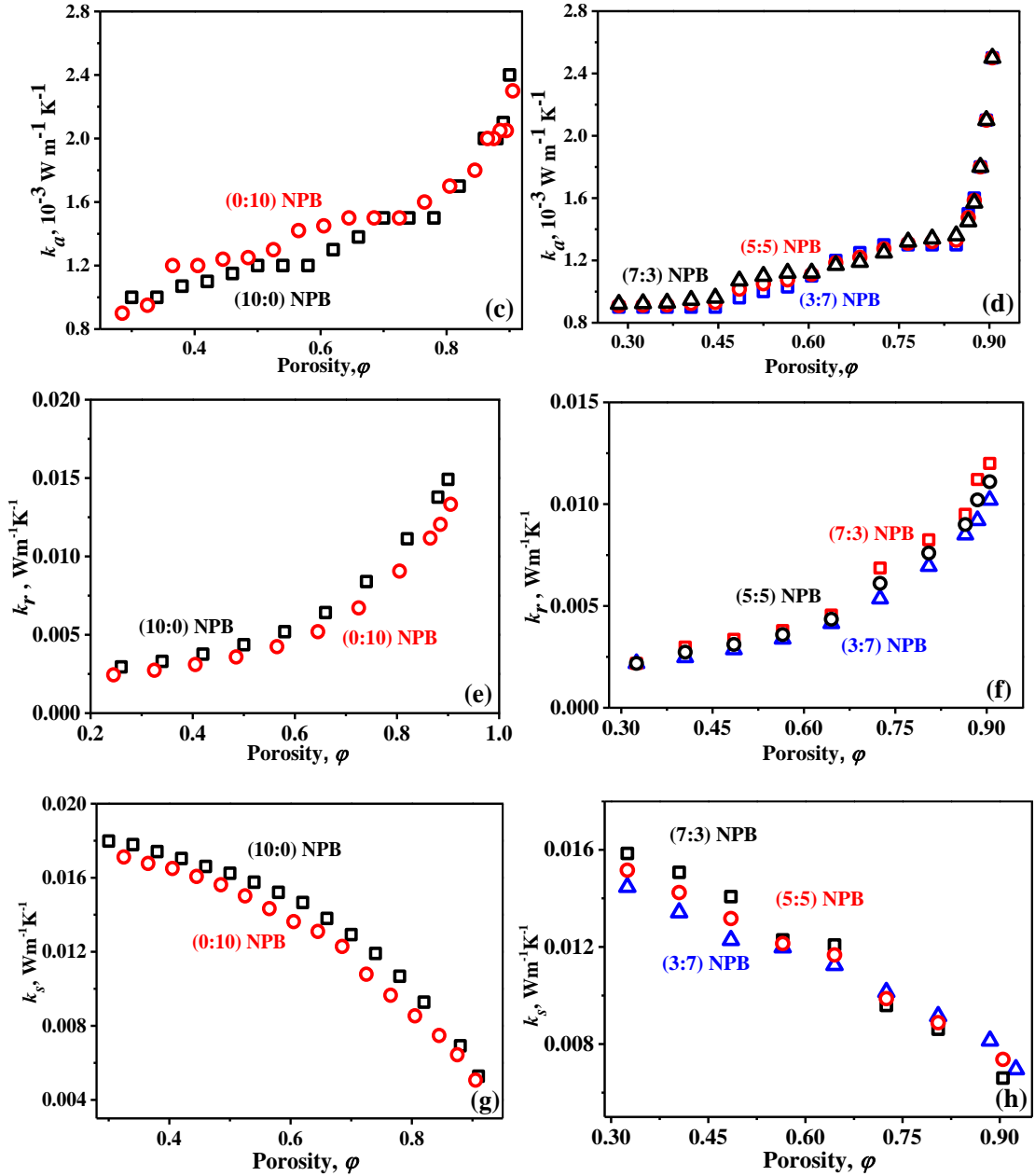

*Fig. 6 $k_v$, $k_a$ and $k_r$ of p-NPB and h-NPB: (a) and (b) show the $k_v$ of p-NPB and h-NPB; (c) and (d) show the $k_a$ of p-NPB and h-NPB; (e) and (f) show the $k_r$ of p-NPB and h-NPB;(g) and (h) show the $k_s$ of p-NPB and h-NPB.*

## 3.3 Mechanism of ultra-low thermal conductivity of NPB

The (0:10) p-NPB, (10:0) p-NPB and (3:7) h-NPB is selected to analyze the mechanism of the ultra-low $k_e$ of NPBs in this part. The values of $k_s$ and $k_r$ are shown in Fig. 7 (a), and the $k_e$ calculated by $k_s+k_r$ is shown in Fig. 7 (b). With the increasing $\varphi$, the $k_e$ firstly decreases and then increases in Fig. 7. The decrease of $k_e$ can be explained by the domination role of the $k_s$, while the increase of $k_e$ at $0.53 < \varphi$ for h-NPB ($0.72 < \varphi$ for p-NPB) is result from the domination role played by the radiation heat transfer. Comparing Fig. 7(a) and (b), we can conclude that the ultralow $k_e$ is determined by the summation of the $k_s$ and the $k_r$ at the $\varphi$ where domination role changes from $k_s$ to $k_r$.

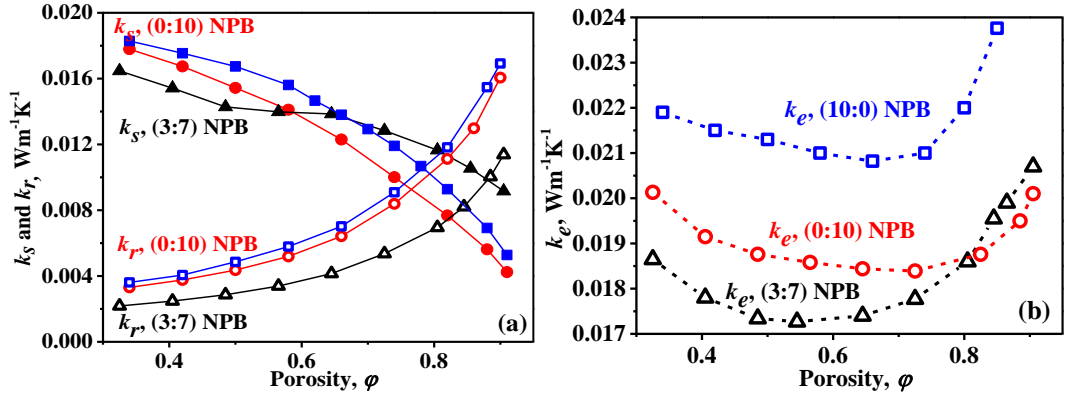

*Fig. 7 Thermal conductivities: (a) $k_s$ and $k_r$ in p-NPBs and h-NPBs; (b) $k_e$ in p-NPBs and h-NPBs;*

Additionally, the $k_e$ of (3:7) h-NPB is lower than that of (0:10) p-NPB and (0:10) p-NPB as illustrated in Fig. 7 (b). To explain the further reduced $k_e$ of h-NPBs, the pore size distribution and the surface area of (0:10) p-NPB, (10:0) p-NPB and (3:7) h-NPB are characterized by the mercury intrusion method with a commercial device (ASIQMU002110-6, Quantachrome Instruments, USA). Results are shown in Figs. 8(a) and (b) and Fig. 9 respectively. Compared with p-NPB, the h-NPB has much smaller pore size and surface area, which indicates that the interfacial contact area between nanoparticles in h-NPBs is larger than that in p-NPBs. The enhanced thermal contact resistance caused by the enlarged interfacial contact area is responsible for the lower $k_e$ of h-NPB than that of p-NPB.

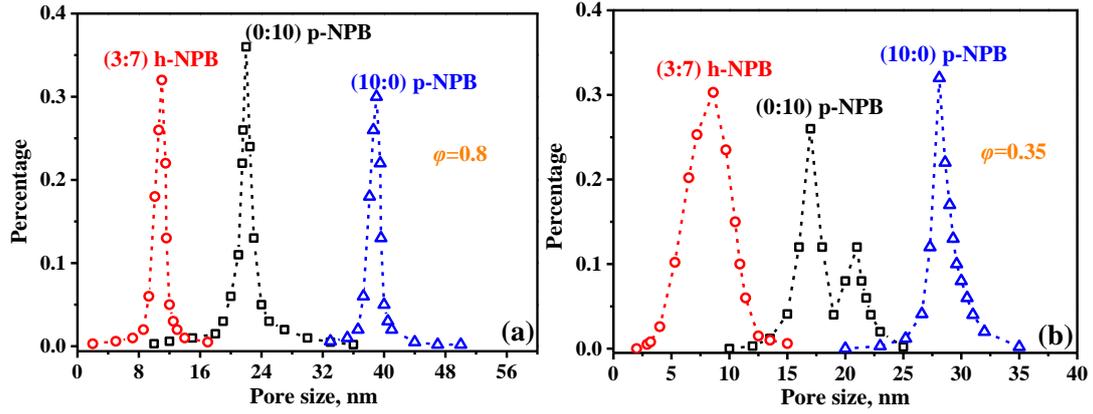

*Fig. 8 Pore size distribution of p-NPBs and h-NPBs: (a) at $\varphi=0.8$; (b) at $\varphi=0.35$.*

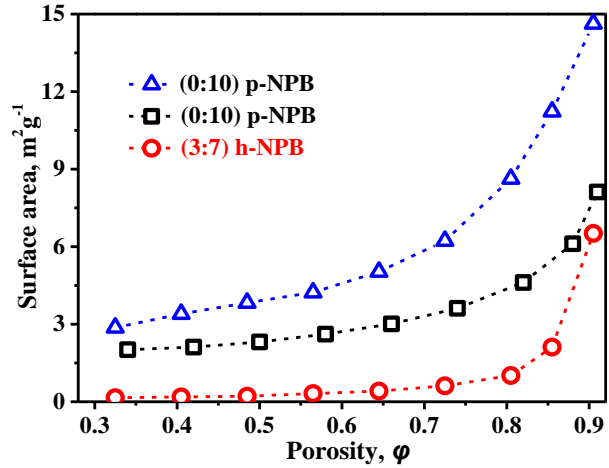

*Fig. 9 $\varphi$-dependence surface area of (0:10) p-NPB, (10:0) p-NPB and (3:7) h-NPB.*

It is worth noticing that the $k_e$ of all samples prepared in this work could be lower than that of the free air (about 0.026 Wm⁻¹K⁻¹ [29]), due to the low $k_s$ attributed to the large thermal contact resistance, the neglegible convection heat transfer of air in nanopores and the small contribution of $k_r$. To further reduce the $k_e$, hybriding smaller nanoparticles may be better, and some methods for increasing the thermal contact resistance should also be developed.

## 4. Conclusion

In this paper, the p-NPBs and the h-NPBs are prepared with a cold-pressing method. The $k_e$ is measured by the hot-wire method and the transient heat source method, and the $k_r$ is calculated by the diffusion approximation model based on the measurement of spectral transmittance ($\tau_\lambda$) observed with a Fourier transform infrared spectrometer. The influence of air on the $k_e$ of a NPB is estimated by subtracting $k_e$ measured at the atmosphere pressure by that measured at a vacuum environment. Results show that the $k_e$ is determined by $k_s$ and $k_r$. With increasing $\varphi$, the $k_e$ of a NPB will firstly decrease because of a decrease of $k_s$, and then increase when radiation heat transfer dominates. The minimum $k_e$ occurs at the $\varphi$ where domination role changes from $k_s$ to $k_r$.

It is worth noticing that the lowest $k_e$ of h-NPB can be as lower as 0.018 W m$^{-1}$ K$^{-1}$. This quite low $k_e$ is due to the quite low $k_s$ caused by the large thermal contact resistance, the negligible $k_a$, and the small $k_r$. To further reduce the $k_e$, hybriding smaller nanoparticles may be better, and some methods for increasing the thermal contact resistance or reducing the radiation heat transfer should be further developed in the future. This study is expected to supply some information for thermal insulation material design, and also to provide some physical insights into the $k$ of a hybrid nanoparticle composite.

**Appendix A: Radiative thermal conductivity.**
The optical thickness is defined as the extinction coefficient times the physical thickness of the material. In practical application, the optical thickness of most insulating materials is always much larger than 1, so the diffusion approximation model can be used, and the radiation thermal conductivity can be written as, [30,31]

$$k_r = \frac{16\sigma T^3}{3E_R},\qquad (A1)$$

where $k_r$ is the radiative thermal conductivity, $\sigma$ is the Stefan–Boltzmann constant, $E_R$ is the Roseland extinction coefficient which can be calculated by,

$$E_R^{-1} = \left[\int_0^\infty \frac{1}{E_\lambda}\frac{dI_\lambda(T)}{dT}d\lambda\right]\left[\int_0^\infty \frac{dI_\lambda(T)}{dT}d\lambda\right]^{-1},\qquad (A2)$$

where $E_\lambda$ is the spectral extinction coefficient, $I_\lambda$ is the spectral radiative intensity of a black body and $\lambda$ is the wavelength of the radiation. When a spectral radiative intensity $I_\lambda$ is incident on a volume element of NPB with thickness $dl$, considering the absorption and scattering, the reduced $I_\lambda$ can be expressed as,

$$dI_\lambda = -E_\lambda I_\lambda dl. \tag{A3}$$

According to the Lambert-Bell law, the relationship between the spectral extinction coefficient and the radiative intensity can be written as,

$$\int_{I_\lambda(0)}^{I_\lambda(s)} \frac{dI_\lambda}{I_\lambda} = \int_0^S E_\lambda(S)dl. \tag{A4}$$

Concerning the NPB is homogeneous and isotropic, the Eq. (S4) can be simplified to [32]

$$E_\lambda = -\frac{\ln(\tau_\lambda)}{L}, \tag{A5}$$

where $L$ is the thickness of the sample and $\tau_\lambda$ is the spectral transmittance measured with a Fourier transform infrared spectrometer (VERTEX 80v Germany Bruker). Substituting Eqs. (A2), (A3) and (A5) into (A1), the radiative thermal conductivity $k_r$ can be calculated. The experimental results of the spectral transmittance for different NPBs are shown in Fig. (A1), and the calculated $k_r$ is shown in Fig. 6 (e) and (f).

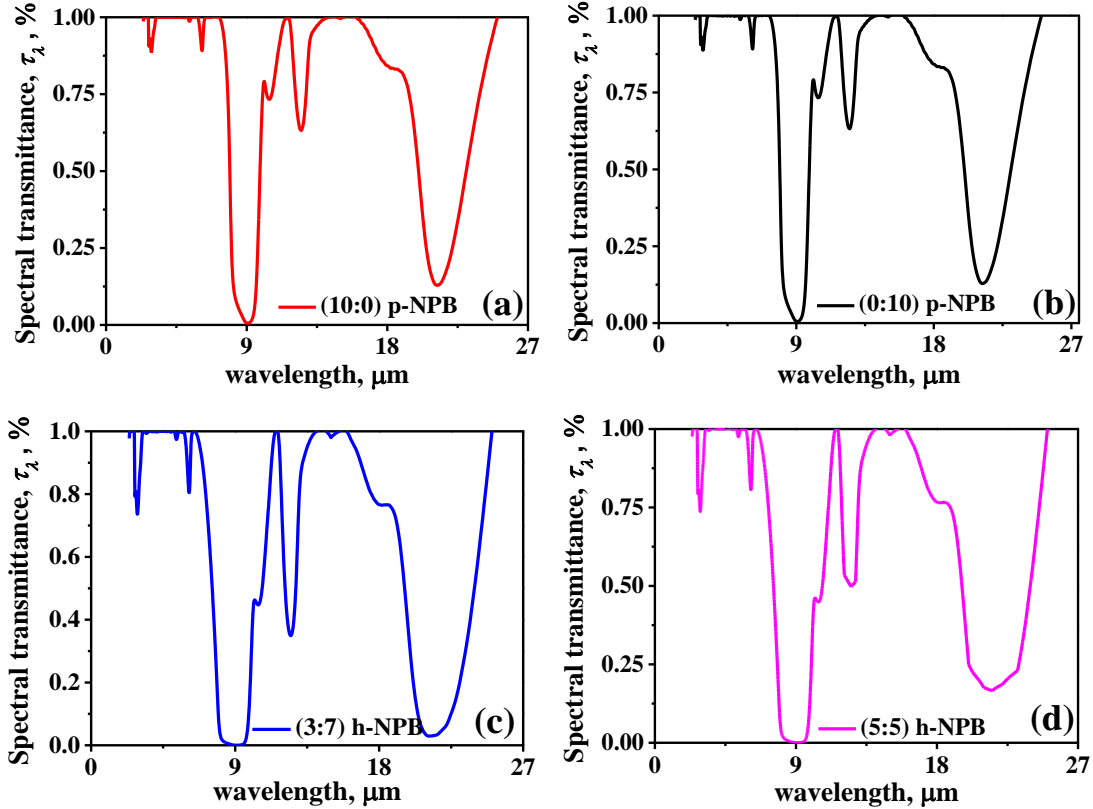

*Fig. A1 Spectral transmittance in NPBs at $\varphi = 0.6$: (a) (10:0) p-NPB; (b) (0:10) p-NPB; (c) (3:7) h-NPB; (d) (5:5) h-NPB.*

**Acknowledgements**

This work is supported by the Fundamental Research Funds for the Central Universities (2019ZDPY06).